\documentclass[a4paper,times, 5p, authoryear]{elsarticle}
\usepackage{color}
\usepackage{url}
\newcommand{\revric}[1]{\textcolor{black}{#1}}
\newcommand{\revyan}[1]{\textcolor{black}{#1}}
\newcommand{\revP}[1]{\textcolor{black}{#1}}
\newcommand{\revPbis}[1]{\textcolor{black}{#1}}

\begin{document}

\begin{frontmatter}
\title{Effect of contact location on the crushing strength of aggregates}
\author[ifsttar]{Riccardo Artoni}
\author[ifsttar]{Aur\'elien Neveu}
\author[ifsttar]{Yannick Descantes}
\author[ifsttar]{Patrick Richard\corref{cor1}}
\ead{patrick.richard@ifsttar.fr}
\cortext[cor1]{Corresponding author}
\address[ifsttar]{IFSTTAR, MAST, GPEM, F-44340 Bouguenais, France}
\date{\today}
\begin{abstract}
This work deals with the effect of the contact location distribution on the crushing of granular materials.
At first, a simple drop weight experiment was designed in order to study the effect of the location of three contact edges on the fracture pattern and the strength of a model cylindrical particle. The sample was placed on two bottom contact edges symmetrically distributed with reference to the vertical symmetry plane of the particle and subjected to an impact at the top. Angle $\alpha$ between the plane connecting a bottom contact edge to the centerline of the cylinder and a vertical plane was varied. The energy required to fracture the particle was shown to be an increasing function of angle $\alpha$. Peculiar crack patterns were also observed.
Then, we present a discrete model of grain fracture based on the work of~\cite{neveu2016} and employ it for a numerical analysis of the problem. The cylindrical particle is discretized by means of a space filling {Vorono\"\i} tessellation, and submitted to a  compression test for different values of angle $\alpha$.
In agreement with experiments, simulations predict a strong effect of the contact orientation on the strength of the particle as well as similar fracture patterns. The effect of the number of contacts is also explored and the importance of a potential pre-load is emphasized.
\revPbis{We show that the fracture pattern: (i) is diametrical in case of diametrically opposed edges, (ii) has an  inverted Y-shape in the case of three or four edges. Interestingly, in the latter case, if one of the lateral edges is slightly shifted, the fracture initiates and even propagates diametrically. Furthermore, the particle strength increases with the number of contacts.}
\end{abstract}
\end{frontmatter}
\section{Introduction}
{Studying the fragmentation of cohesive materials is of importance for a wide range of natural and industrial processes. 
{As an example, avalanches and landslides are often modelled  as cohesive granular materials~\citep{Langlois_JGeoPhysResEarthSc_2015} and consequently, understanding how such geophysical events are triggered and how far they can flow as well as quantifying the damages suffered requires to understand their failure mechanisms.
Similarly, Asteroids can be seen as collections of grains interacting through, among others, cohesive forces~\citep{Richardson_PlanetarySpaceScience_2009}. It has been shown that their properties (\textit{e.g.} shape) may strongly depend on the cohesion forces and on their impact on the internal stresses.} \\
{From an industrial point of view, crushing occurs in a large number of processes.
For instance, in the production of aggregates, rock blocks are crushed and the resulting fragments are required to meet high standards namely in terms of size and
shape. Successive crushing steps are usually carried out to achieve the requested aggregate characteristics, leading to a waste of
good quality raw materials and a high energy cost that could both be mitigated upon improving the crushing efficiency.}\\

An important aspect of the crushing process is that loads are applied on a collection of particles, and therefore stresses are transmitted through particle contacts. 
In this perspective, some authors have proposed failure criteria for tensile~\citep{Tsoungui1999} and plane shear~\citep{BenNun2009} fracture modes taking into account the effect of the
coordination number by lumping randomly distributed contact forces in a set of 2 main forces acting along force network eigenvectors.
However, the effect of anisotropy of the contacts' location on the strength of the particle has not been
deeply investigated.}
{ 
\cite{Todisco2015,Todisco_SoilsFund_2017} constructed an experimental device to study multicontact
crushing of  sand or limestone particles between a set of
steel balls or particles glued to the loading frame. They show that a higher coordination number
leads to a lower probability of crushing.
Using an original experimental device, \cite{Salami_EJECE_2017}, confirm that the number of contacts and their positions
 play an important role in the fragmentation of an individual
particle. 
They conclude the existence of two types of 
cracks each corresponding to a different crack mode and dependent
on the contact arrangement. 
For the first one, the cracks appeared to originate near the contacts and propagated towards
the particle centre. For the second one, the cracks also originate from the contacts but propagate towards the farthest load bearing
contact.\\
Here, we apply a discrete-element numerical model --previously used to simulate cemented materials~\citep{neveu2016}-- to study the fracture of voidless materials and probe this method by means of experiments.
Our middle term objective is to simulate the crushing of aggregates by using a method valid for any type of materials, \textit{i.e.} cemented materials like sandstone or voidless materials like marble.  For this reason, Discrete Element Methods is a natural choice.\\ 
%
In this paper, we first examine from simple drop weight crushing tests how the contact location distribution on a cylindrical particle affects the particle strength and the fracture pattern.
Then we employ a discrete model of cohesive material to study the effect of contact location anisotropy on the shear strength. First the model is introduced, then numerical simulations of compression of cylindrical particles are presented. 

\section{Experiments :  drop weight tests}
\begin{figure}[htbp]
\centering
\includegraphics[width=0.48\textwidth]{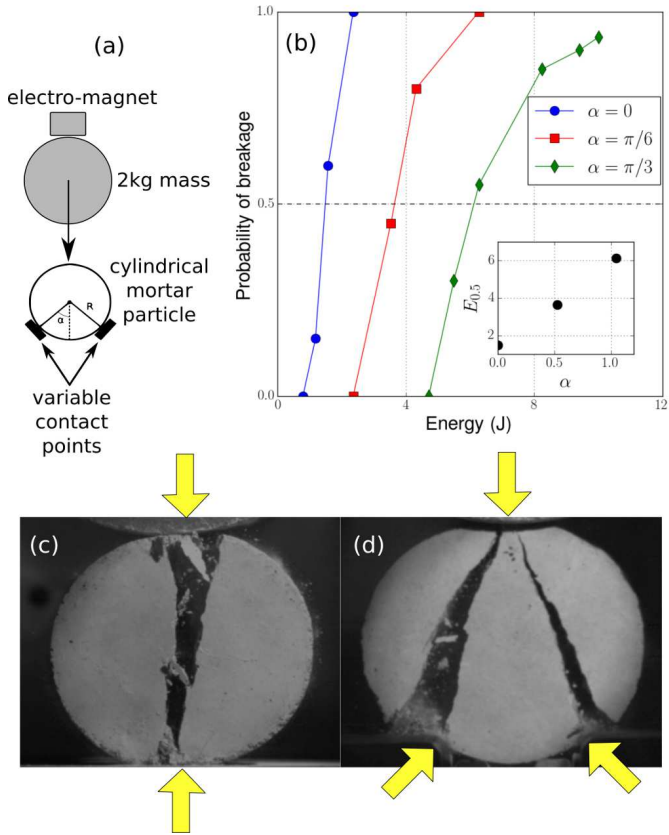}
\caption{(a) Sketch of the experimental setup. A 2kg mass \revyan{suspended to an electro-magnet} was dropped from a variable height on a  cylindrical mortar sample {having two contacts characterized by \revyan{angle} $\alpha$}; {(b) For a fixed energy value, the probability of breakage decreases when $\alpha$ increases.
Fracture patterns of the sample from high speed video recordings are clearly influenced by the number of contacts : 2, \revyan{when} $\alpha=0$ for (c)  and 3 with  $\alpha=\pi/6$ for (d)}.}
\label{cucu}       
\end{figure}

In order to 
{investigate} the effect of contact location distribution on particle fracture and strength, a simple experiment was performed. Cylindrical mortar samples ($D=50$ mm, $H=25 $ mm) were tested using a drop weight setup which allowed varying the 
{location} of contacts at the bottom (see Fig.~\ref{cucu}a). 
In the following we will 
{parameterize the location} of these contacts by  $\alpha$, the angle 
between the plane connecting a bottom contact edge to the centerline of the cylinder and a vertical plane.
A $2~\mbox{ kg}$ cylindrical steel mass ($D=100$ mm, $H=30$ mm) held by an electro-magnet was dropped from variable height. The shock energy transmitted to the sample was assessed from the dropping height whereas a high speed video camera recorded the sample behaviour. {The probability of breaking the sample is depicted in Figure \ref{cucu}b as a function of input energy (drop height) for different values of angle $\alpha$: $\alpha=0$, $\pi/6$ and $\pi/3$. For each energy value, $20--30$ tests were performed.} In agreement with the conclusion of \cite{Todisco2015,Todisco_SoilsFund_2017}, figure~\ref{cucu}b shows that the energy required to crush the particle is an increasing function of angle $\alpha$, at least in the range considered. This is even clearer when looking at the inset in Figure \ref{cucu}b, where the interpolated shock energy required to achieve $50\%$ chance of breaking a sample is plotted as a function of angle $\alpha$. It is obvious that the effect of $\alpha$ is very strong, given that the energy required for crushing at $\alpha=\pi/3$ is approximately four times larger than that at $\alpha=0$. Figure \ref{cucu}b also shows that the energy required to break the sample is more dispersed and  also that it increases with $\alpha$. This is confirmed by fitting the probability of breakage by a Weibull law: we find that the Weibull exponent, $m_W$, decreases with increasing $\alpha$ : 
$m_W\approx 6$ for $\alpha=0$, $m_W\approx 4.9$ for $\alpha=\pi/6$ and $m_W\approx 3.2$ for $\alpha=\pi/3$. 

The fracture pattern of the sample for two contact configurations ($\alpha=0$ and $\alpha=\pi/6$) is also shown in Fig.~\ref{cucu}c-d.
It is clear that in the $\alpha=0$ case, a classical diametral fracture pattern is observed, with the development of a vertical crack between the two contact edges. On the other hand, when three contact edges are present, two cracks develop between the impact edge and the two bottom contact edges. 

From these simple experiments and in agreement with the literature~\citep{Todisco2015,Todisco_SoilsFund_2017,Salami_EJECE_2017}, we can deduce that both particle strength and fracture pattern are strongly influenced by the contact configuration. 
%
%
%
In order to investigate the micromechanics of particle fracture, we introduce in the following a discrete numerical model.

\section{Discrete numerical model}

Numerical simulations using the well-known Discrete Element Methods (DEM) have already been successfully used to describe the breakage behavior of granular materials~\citep{Potyondy2004,Weerasekara2013,Andre2012}. Different approaches have been proposed. For example, \citep{Potyondy2004} modeled a rock piece as a collection of spheres or disks with cohesive, elastic-brittle bonds at interparticle contacts, and solved the mechanics in the framework of molecular dynamics. On the contrary,~\citep{topin2007}  modeled a cemented material by means of a subparticle lattice discretization, that is a triangular network of springs with stiffness and strength properties depending on the material.
\cite{Nguyen2015} proposed instead a model based on contact dynamics in which there is no elasticity in cohesive interactions, which are represented as perfectly rigid bonds with a threshold value on the tensile and shear stress above which  the interaction is broken (yield criterion). Other models~\citep{Riviere_TribLetters_2015} rely on cohesive forces at particle contacts which possess complex yielding criteria coming from the framework of cohesive zone models~\citep{Rouas_CompMethApplMechEng_1999}. Other approaches treat contacts and cohesive interactions by superimposing a lattice discretization to a molecular dynamics scheme~\citep{Daddetta2002}.

In a recent work,~\cite{neveu2016}, we introduced a cohesive interaction model for Discrete Element Methods which allows to model cohesion between contacting and non-contacting particles, suitable for any kind of particle shape. The model treats independently contacts and cohesive interactions.  A grain is represented by a collection of particles with bonds to model inter-particles cohesion.

Cohesion is set inside the modeled material by applying forces which
oppose relative motion between particles. This relative displacement
is computed between two cohesion points, each located \revyan{inside} one of the
two interacting particles. Cohesive interactions
consist of a spring and a damper for both normal and tangential components. As cohesion
is treated separately from contact, this allows to apply
cohesive forces even if particles are not touching each other, whatever
their shape.
More than one cohesive interaction may be set between
{two} interacting particles, if necessary, to resist rotation, bending and torsion. 

In~\citep{neveu2016}, we studied the behavior of the model 
with respect to simple compression tests in two dimensions of a cemented material. In that case, cohesive interactions were meant to mimic the presence of the cement paste.

In this work we study the behavior of a solid grain discretized by a {Vorono\"\i} tessellation~\citep{Voronoi1907}
which is classically used to fill space from a geometrical distribution of points or spheres~\citep{Hanson_JStatPhys_1983,Richard1999,Richard1998}.
Interparticle contacts are modeled with unilateral contact laws in the framework of non smooth contact dynamics~\citep{jean99}. In particular, collisions are inelastic with a non-interpenetration constraint and Coulomb friction holds for the tangential component in long lasting contacts.
\revric{As regards the modeling of cohesion, one cohesive interaction is set between every pair of {Vorono\"\i} cells which share a face (Fig.~\ref{fig:sketch}). Cohesive interactions are modeled as elastic-viscous links with a brittle failure criterion, and the seeds of the Vorono\"\i\, tesselation are chosen as cohesion points. This ensures that in the initial undeformed state, the vector linking the cohesion points is perpendicular to the surface shared by the two cells.
In the reference frame of the shared face plane taken as the 
cohesive interaction plane, the  cohesive force acting on the cohesion point of particle $p$, $\vec x_p^{\,c}$, by the action of cohesion point $\vec x_q^{\,c}$ belonging to particle $q$  writes: 
\begin{equation}
\vec{F_{pq}}=F_{pq}^{n} {\vec n_{pq}}+F_{pq}^{t} {\vec t_{pq}}+F_{pq}^{s} {\vec s_{pq}},
\end{equation}
with the normal ($n$) and tangential ($t,s$) force components given by:
\begin{equation}
F_{pq}^{i}=-k_{pq}^{i}\Delta_{pq}^{i}-\eta_{pq}^{i}U_{pq}^{i}, \mbox{ with } i=n,t,s
\end{equation}
where $\Delta_{pq}^{i}$ are the components of the relative displacements in the cohesive interaction
plane reference frame, and the $k_{pq}^{i}$ are the corresponding stiffnesses. Energy dissipation is accounted for through viscous damping\revyan{, taken critical in order to avoid oscillations}: $U_{pq}^{i}$ denotes the relative velocity of the interaction points along direction \emph{i} and $\eta_{pq}^i$ the corresponding damping coefficients. 
In particular, the relative displacements are given by 
\begin{eqnarray}
\Delta_{pq}^{n}=(\vec x_q^{\,c}-\vec x_p^{\,c})\cdot \vec n_{pq} - l_{pq}^{c,0}\\
\Delta_{pq}^{t}=(\vec x_q^{\,c}-\vec x_p^{\,c})\cdot \vec t_{pq} \\
\Delta_{pq}^{s}=(\vec x_q^{\,c}-\vec x_p^{\,c})\cdot \vec s_{pq} 
\end{eqnarray}
where $ l_{pq}^{c,0}$ is the initial distance between cohesion points, and the relative velocities by projecting the relative velocity vector 
\begin{equation}
\vec U_{pq}=\vec v_q + \vec \omega_q \times (\vec x_q^{\,c}-\vec x_q^{\,b}) -\vec v_p - \vec \omega_p \times (\vec x_p^{\,c}-\vec x_p^{\,b})
\end{equation}
on the cohesive interaction reference frame. This last equation simply \revyan{reflects} the fact that, due to the placement of the cohesion points on the seeds of the Vorono\"\i\, cells $\vec x_p^{\,c}$ and $\vec x_q^{\,c}$, which are in principle different from the centers of mass of the particles $\vec x_p^{\,b}$ and $\vec x_q^{\,b}$ to which motions are referred,  the relative velocity between cohesion points receives contributions from both the translation and the rotation of the particles.
In order to recover a local Hookean behavior, the stiffnesses $k_{pq}^{i}$ are given by
\begin{equation}
k_{pq}^{i}=\frac{k_{pq}^{i*} S_{pq}}{l_{pq}^{c,0}},\label{eq:kn}
\end{equation}
where $k_{pq}^{i*}$ denotes a cohesive (normal or tangential) stress scale, and  $S_{pq}$ is the cohesive interaction surface, taken here equal to the area of the face shared by the {Vorono\"\i} cells in contact. Since the cohesive interaction surface
$S_{pq}$ and length $l_{pq}^{c,0}$ differ for each pair of particles, a microscopic
stiffness heterogeneity is introduced.}

A simple brittle, irreversible breakage criterion was assumed for the cohesive interactions: a cohesive interaction breaks in tension or shear when the relative displacement reaches one of the following threshold values:
\revric{\begin{equation}
\Delta_{pq}^{i,max}=\frac{l_{pq}^{c,0}}{k_{pq}i^{i*}}\,.\,\sigma_{i,r},\label{eq:delta_max}
\end{equation}}

with $\sigma_{i,r}$  the microscopic (direction-dependent) strength which is an input parameter of the model, taken identical for all the interactions. 
{The choice of this criterion is motivated by its simplicity.}

\begin{figure}
\begin{center}
\includegraphics[width=0.4\textwidth]{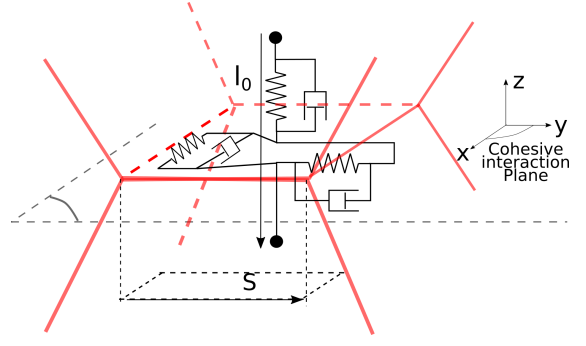}
\end{center}
\caption{(a) \revyan{3D} sketch of the cohesive interaction set between the centers of two {Vorono\"\i} cells, with a spring-dashpot system along the normal and tangential components.
The initial length of a cohesive interaction $l_0$ is the distance between the two centers of the cells and $S$ is the cohesive interaction surface.}
\label{fig:sketch}       
\end{figure}

\begin{figure}[htbp]
\begin{center}
\includegraphics[width=0.55\columnwidth]{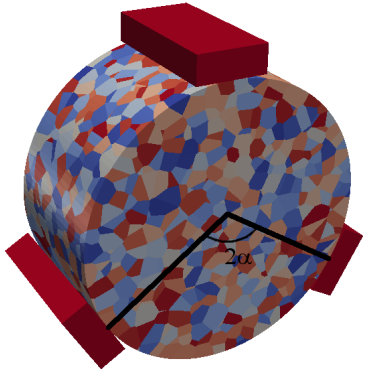} 
\end{center}
\caption{Perspective view of a 3d {Vorono\"\i} tessellation generated with Neper~\citep{neper} (different colors for visualization purposes only). {The number of cells is equal to $5000$. The angle between the two lateral contacts is $2\alpha = 2\pi/3$.}}
\label{sketchb}       
\end{figure}

The model described in the previous section has been implemented in the Contact Dynamics framework LMGC90~\citep{renouf04}. It is applied here to study the crushing of a grain subjected to a multi-edges loading. 

The simulated grains are cylinders of diameter $d$ and thickness $0.5 d$. They are built by generating a
3d {Vorono\"\i} tessellation by means of the Neper software~\citep{neper}.
The envelope of the cylinder parallel to its axis is discretized in 290 identical rectangles, the cylinder basis being therefore a regular polygon with 290 edges.
%
%
%
The {Vorono\"\i} tessellation leads to a subdivision of the studied domain in smaller and simpler subdomains which have the advantage to be disordered and not \revyan{degenerated} \textit{i.e.} each vertex of the tessellation is shared by exactly three edges (see Fig.~\ref{sketchb}). 
Since, in our model, the failure of the sample is a consequence of the breakage of cohesive bonds between interacting cells, it is important to subdivide the domain finely.
{In other words, the choice of the discretization level is a key parameter since it gives the possible fracture paths. Obviously, the larger the number of cells used to discretize the sample, the better the description. Yet to save computation time we carried out simulations with various numbers of cells, $N$, to test the influence of the discretization level and see if the mechanical response of the grain becomes independent of $N$ above a threshold value.  
\begin{figure}[htbp]
\centering
\includegraphics[width=0.85\columnwidth]{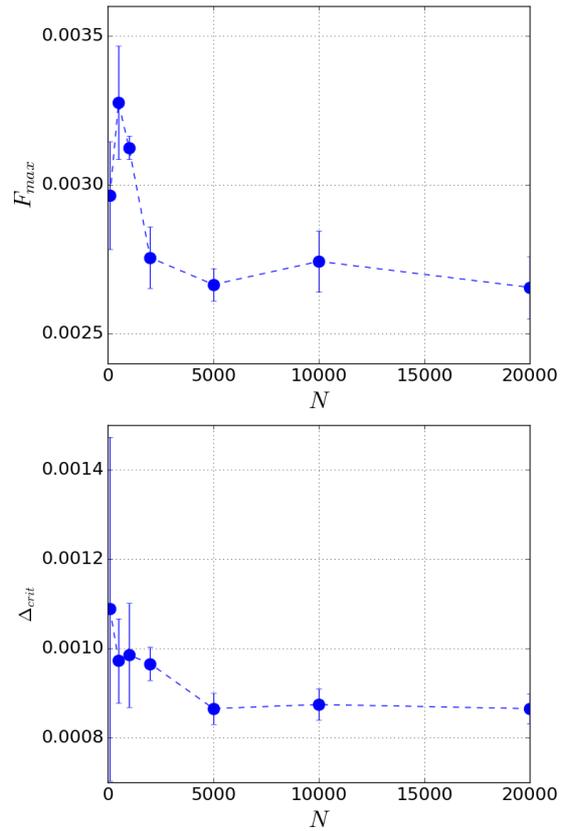}\\
\caption{(a) Force peak, (b) critical displacement for different values of the discretization $N$, for diametral compression, case A (tangential elasticity, no friction).
The forces and the lengths \revyan{are} given in nondimensional quantities by using respectively 
the following normalization parameters: \revric{$k^*d^2$} and $d$.}
\label{mech2}       
\end{figure}
Figure~\ref{mech2} shows, for a diametral compression,  the evolution of $F_{max}$, the maximum force measured during  compression and that of $\Delta_{crit}$, the corresponding displacement with $N$,  the number of cells used to tessellate the grain.
As expected $N$ has an effect on the mechanical response of the grain, and on the scattering of data. Asymptotic values are obtained for sufficiently large $N$. Unless otherwise stated, the analyses that follow correspond to the value $N=5000$ which gives a good balance between computational cost, independence from $N$ and data dispersion.} 


Two, three or four walls are placed in contact with the sample. In \revyan{each} case, one of them is in contact with the top of the sample (top wall). It is moved at constant velocity to load the grain.  In the case of three contacts, as in experiments, two fixed lateral walls are positioned symmetrically, with their normal pointing towards the centerline of the cylinder. These positions are characterized by the half angle $\alpha$ between planes connecting each lateral contact to the centerline of the cylinder (as defined in Fig. \ref{cucu}a). $\alpha$ is varied between $0$ and $\pi/3$.
In the case of four contacts,  a fourth wall,  diametrically opposed to the top wall is added. The corresponding contact  is called bottom-contact. 
Three repetitions were made for each angle. Simulations were performed without gravity. The stress scale is given by the microscopic cohesive stress scale $k^*$, the mass scale by the mass of the cylinder $m$, the length scale by the cylinder diameter $d$, and the time scale by $\sqrt{{m}/{(k^* d})}$. 
In order to study the effect of the model chosen for the tangential interactions, two cases were considered: 
\begin{itemize}
\item Case A:  the tangential and normal cohesive stress scales are equal: 
$k_{n}^{*}=k_t^*=k_s^* = k^*,$
 and so are the microscopic strengths $\sigma_{n,r}=\sigma_{t,r} =\sigma_{s,r}=\sigma_{r}; $ there is no friction between particles, $\mu_{pp}=0$, but friction with the walls, $\mu_{pw}=0.3$.
\item Case B:  there is no tangential elasticity, $k_{n}^{*}=k^*; k_t^*=k_s^* = 0$, and no tangential breakage; there is friction between grains and between particles and walls, $\mu_{pp}=\mu_{pw}=0.3$. In other words, the tangential component of the cohesion force is set to zero.
 \end{itemize}
The cylinder was loaded by vertical translation of the top wall at constant 
velocity  chosen of the same order of magnitude as in the impact test experiments, $V_c=1\cdot10^{-4} \sqrt{{m d}/{k^*}}$. 
\revP{Note however that, unlike in experiments, the impact is modeled as a constant velocity in the numerical simulations.}
The breakage stress was kept fixed at $\sigma_{r}=5\cdot 10^{-3} k^*$.

\section{Numerical results}
\subsection{Mechanical behavior for $\alpha = 0$.}
The fracture pattern obtained in simulations for $\alpha=0$ is similar to that obtained experimentally under the same conditions: \revric{a vertical fracture is found, which is initiated by the breakage of \revyan{a} few cohesive interactions located  within the sample close to the symmetry plane. Then, cohesive links continue to progressively break around the symmetry plane until a crack percolates}.
Sample force-displacement curves obtained from simulations \revyan{are} plotted in Fig. \ref{mech}. We can see that each curve displays a linear, elastic part, then reaches a maximum and decreases gently, suggesting the occurrence of progressive damage of the grain. This behavior is confirmed by the evolution of the number of broken cohesive interactions: for small displacements no breakage of the cohesive interactions occurs, then the sample is progressively damaged. Thus, the material  experiences quasibrittle fracture, which is typical of microstructured materials such as concrete. This point is particularly interesting: keeping in mind that cohesive interactions are perfectly brittle,
 a quasi brittle macroscopic behavior can be obtained by the interplay of cohesion heterogeneity (which makes some cohesive links break at low values of the wall displacement) and geometrical frustration (displacement of the bodies, and thus fracture propagation, is hindered by the space filling nature of the {Vorono\"\i} packing). Figure~\ref{mech} allows to perform a first evaluation of the effect of assumptions made to describe the tangential components of contacts and interactions. Clearly, case B (no tangential elasticity, interparticle friction) displays a higher value of the force maximum (apparent strength) with respect to case A (no interparticle friction, tangential elasticity). Concerning the post-peak behavior, the decay of the force is slower for case B, and so is the evolution of the number of broken cohesive interactions. Therefore the substitution of interparticle friction for tangential elasticity increases the quasibrittle nature of the model material.
Given the shape of the force-displacement curves, in the following we will characterize the breakage behavior of the numerical samples by two macroscopic parameters, the value of the force at its maximum $F_{max}$ and the critical displacement corresponding to the force peak, $\Delta_{crit}$.
Note that,  this change in the macroscopic behavior of material, consequence of a modification of the interaction model is similar to what has been observed with the same model in cemented materials~\citep{neveu2016}, in which a loss of brittleness induced by a decrease of the number of cohesion links has been observed.
An important point should be emphasized: the force peak corresponds to the initiation of the crack not to its percolation through the material.  To achieve the latter percolation a displacement of the top wall equal to approximately twice the displacement corresponding to the force peak is typically required. 

\begin{figure}

\includegraphics[width=0.95\columnwidth]{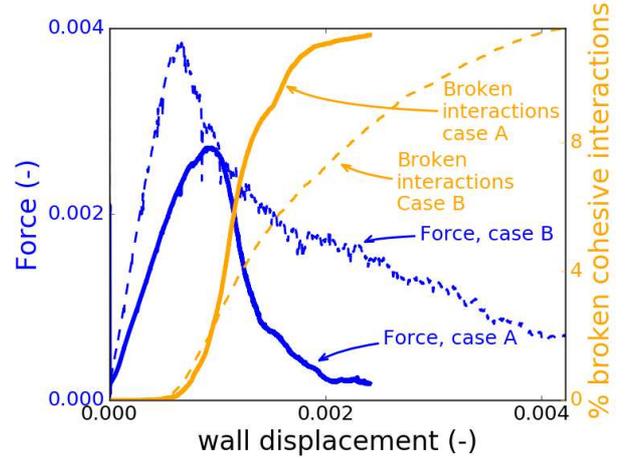}\\

\caption{{The Force-displacement behavior (solid curves) and the evolution of the broken cohesive interactions (dashed curves) for diametral compression (\textit{i.e.} $\alpha=0$), $N=$5000 elements, case A (tangential elasticity, no friction) displays a quasi-brittle nature. Case B (no tangential elasticity, interparticle friction) suggests even more quasibrittle behavior is reinforced (dashed curves).}}
\label{mech}       
\end{figure}

%
%
\subsection{Three contacts: effect of the location}
For both models described above, figure~\ref{mech3} shows the effect of angle $\alpha$ on the breakage parameters of the numerical sample. {As expected, and in agreement with the literature~\citep{Salami_EJECE_2017,Todisco2015,Todisco_SoilsFund_2017} the position of contacts influences the maximum force supported by the material, and therefore its apparent strength. As it was shown  in Figure~\ref{mech}, also friction  contributes to an increase of $F_{max}$. Figure~\ref{mech3} highlights that the effect of friction increases with $\alpha$. This is probably because increasing   angle $\alpha$ leads to an increase in the shear component of the loading (which for $\alpha=0$ is close to $0$ as simple indirect traction dominates) and therefore the tangential components of interactions and contacts have an increasing role for higher $\alpha$. The critical displacement does not display a large variation with $\alpha$, possibly because displacements in our {Vorono\"\i} tessellation made of non-deformable particles concentrate in the fractures, hence they scale with the maximum displacement permitted in each cohesive bond crossed by the fracture. The slight increase of $\Delta_{crit}$ for large $\alpha$ can be explained by \revric{slight squeezing associated by} slip at the lateral walls.}
\begin{figure}

\includegraphics[width=0.85\columnwidth]{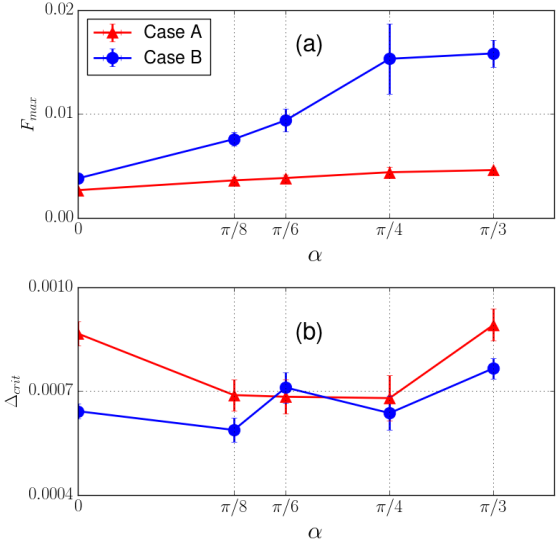}\\

\caption{(a) The force peak increases with angle $\alpha$ whatever the  model  used (red triangles: model A and blue circles model B) and (b) critical displacement for different values of the angle $\alpha$, model A (red triangles), and model B (blue circles). The presence of friction in the  model (\textit{i.e.} model B) increases the values of the maximum force and reduces the variations of the critical displacement.}
\label{mech3}       
\end{figure}


\begin{figure}
\centering
\includegraphics[width=\columnwidth]{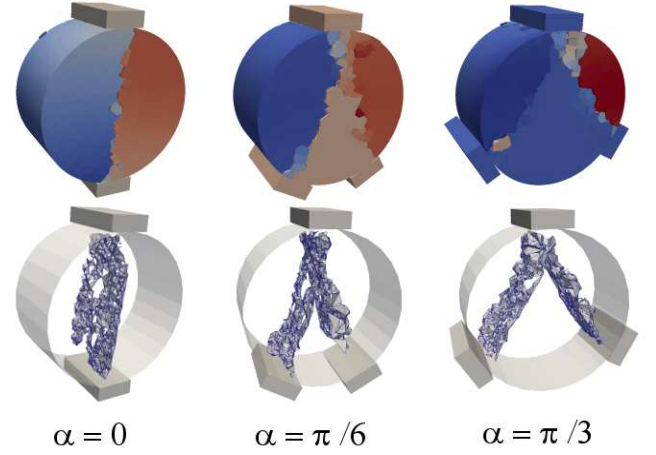}\\
\caption{Displacement fields for the numerical simulations for three values of angle $\alpha$ (top). Colors correspond to particle displacement in the horizontal direction. Corresponding location of broken contacts (bottom). The force model used is case A (elastic).}
\label{pipi}       
\end{figure}

Figure~\ref{pipi} displays \revyan{a representation of} the \revyan{main percolating} fracture patterns obtained for three values of angle $\alpha$: $0$, $\pi/6$ and $\pi/3$. 
\revyan{Each percolating fracture results from a surface reconstruction} algorithm inspired from~\citep{Chaine03} and \cite{Zhao00}. \revyan{This algorithm allows} to approximate fracture surfaces from the cloud of broken cohesive links, each broken link being represented by its center point. 
Basically, this algorithm consists in deforming an oriented initially flat surface made of triangular facets that is embedded in the 3D Delaunay triangulation of the points cloud, until each facet of this surface becomes a Delaunay triangle of the points cloud. Since then, the algorithm has converged to a pseudo-surface consisting of a piecewise linear approximation of the fracture surface. 
\revyan{It should be pointed out that, next to a percolating fracture, a finite size damage zone was also evidenced, which extends roughly over $3/10$ of the cylinder diameter regardless of} \revric{the discretization for $N>5000$. The existence of this zone can be related to the quasi-brittle behavior of our model material. As mentioned above, for $\alpha=0$ a vertical crack is formed.} For  three contact edges, two cracks are obtained between the top wall and the lateral contact edges. 
The cracks initiate   \revric{in the upper part of the sample} and then propagate  downwards until they reach the lateral walls. Similarly to what has been observed in the  case $\alpha=0$, the force peak corresponds to the initiation of the cracks but their propagation goes on beyond the peak displacement and is over before twice this displacement.
The fracture pattern is therefore the same as in the experimental results. The locations of broken force interactions at force peak support this point. The relative dispersion of the locations of broken interactions is related to the discretization. 


We have reported in Fig.~\ref{fig:forces3ptelaspi_6} the evolution of the number of broken cohesive links  as well as the variations of the normal forces at contacts (\textit{i.e} top, bottom and lateral contacts) \textit{versus} strain. Similarly to what has been observed for $\alpha=0$, the forces  display a linear part, reach their maximum and then decrease gently, thus evidencing a progressive damage of the grain. 
The two symmetric lateral contacts experience the same force until the first crack appears leading to a break of symmetry. 

\begin{figure}[htbp]
\centering
\includegraphics[width=\columnwidth]{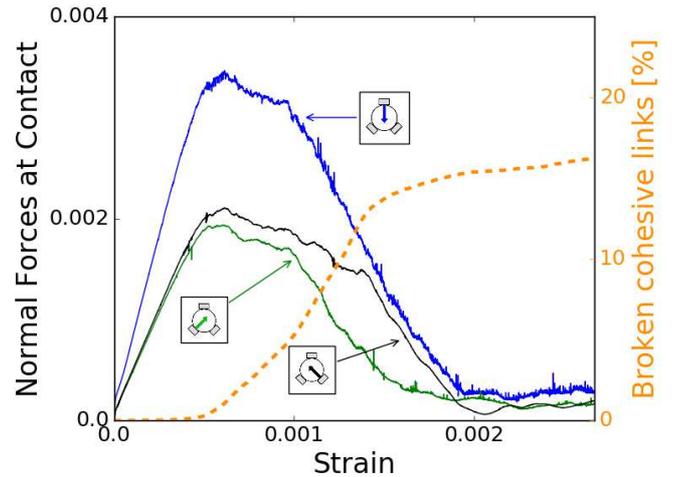}\\
\caption{Evolution with strain of (i) the number of broken cohesive links and (ii) of the normal forces at contacts in the case of elastic interactions (case A) and three contacts. The angle $\alpha$ characterizing the location of the bottom contacts is equal to $\pi/6$.}\label{fig:forces3ptelaspi_6}
\label{fig:forces_3pts_elas_pi_6}       
\end{figure}

\subsection{Four contacts}
To go further in our analysis we also 
study configurations with four contact points.
As mentioned above, for that purpose the four contact configuration we use is inspired by that of~\cite{Salami_EJECE_2017}. Similarly to what has been done before, the wall contacting the sample at its top 
is moved at constant velocity to load the body. A second wall, with a fixed position, is diametrically opposed to the latter wall. The corresponding contact  is called bottom-contact. Two other fixed walls (the corresponding contacts are hereafter called lateral-contacts) are positioned symmetrically at the bottom, with their normal pointing towards the centerline of the cylinder with an angle $\alpha=\pi/4$ (Fig.~\ref{fig:sketch_4points}). 
\begin{figure}[htb]
\begin{center}
\includegraphics*[width=0.55\columnwidth]{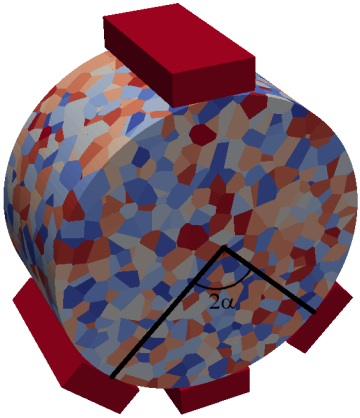}
\caption{Face view of a 3d {Vorono\"\i} tessellation generated with Neper~\citep{neper} 
(different colors for visualization purposes only). The number
of cells is equal to 5000. The angle between the two lateral-contacts is
$2\alpha = 2\pi/4$.}\label{fig:sketch_4points}
\end{center}
\end{figure}
\revP{Note that initially, the three contact walls located below the center of the grain are not loaded hence the corresponding deformations are zero though the contacts are established.}
This latter point is noteworthy since it induces an important difference between the present work and that of ~\cite{Salami_EJECE_2017}
. As we will discuss below, the implications of this preload on the fracture patterns are strong.\\
The simulated particle is made of $N=5000$ {Vorono\"\i}
cells and, similarly to what has been done previously, both elastic (case A) and frictional interactions (case B) between cells will be considered.
\begin{figure}[htbp]
\begin{center}
\includegraphics*[width=0.85\columnwidth]{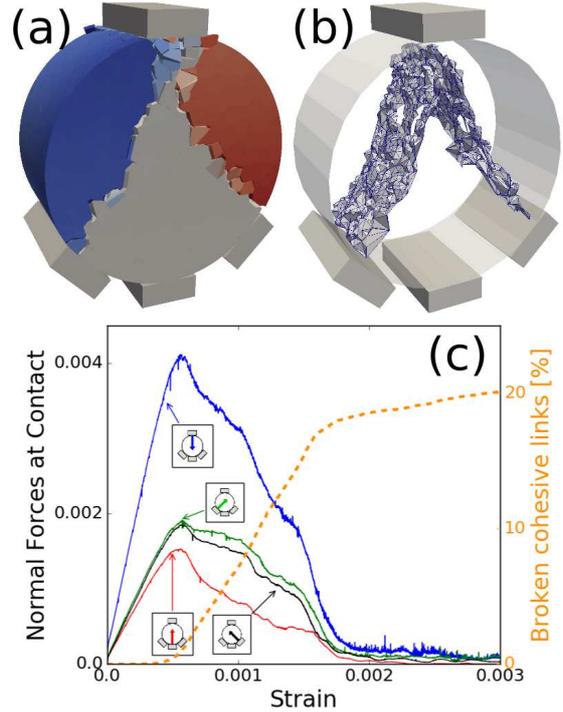}
\caption{Displacement field for the numerical simulations of the four contact configuration in which all the contacts are initially active (a) and the corresponding fracture surface (b).
Evolution with strain of (i) the number of broken cohesive links and (ii) of the normal forces at contacts in the case of elastic interactions \textit{i.e.} case A (c).}
\label{fig:patterns_4pts}
\end{center}
\end{figure}
Interestingly, in such a configuration, and whatever the force model used (case A or case B) no 
central crack appears along the diametric loading axis as illustrated in Figure~\ref{fig:patterns_4pts}(a) and (b) for elastic interactions (case A). On the contrary, cracks,
which are always initiated at the loading \revyan{edges}, propagate towards the closest contact \revyan{edges} \textit{i.e.} the lateral contacts which are located at $\alpha=\pi/4$. 
We have reported in Fig.~\ref{fig:patterns_4pts}c the evolution \textit{versus} strain of (i) the forces at the four contacts (lateral contacts \textit{i.e.} left and right contacts as well as top and bottom contacts) and (ii)  the increase of the number of broken links  during the compression process. Several information deserve to be pointed out. First, the increase of the number of broken cohesive links is rather smooth in agreement with the quasi-brittle behaviour mentioned above. Second, as expected, the evolutions of the forces with the strain deviate from linearity when the first cohesive links are broken. Third, the forces at the lateral contacts are significantly larger than that at the bottom contact, which is consistent with the formation of cracks between the loading \revyan{edge} and the lateral ones. Note that, this remains the same in the case of frictional interaction (case B, not shown) but the differences are less significant. 

Our results are in disagreement with those of \cite{Salami_EJECE_2017}, in which, (i) a systematic presence of a diametral crack has been reported and (ii) secondary cracks originate near one of the lateral contacts, and
propagate towards the farthest load bearing contact (here, the top contact).
In order to explain these differences, let us focus on the way stresses are applied on the particle both in our simulations and in the experiments reported in~\citep{Salami_EJECE_2017}. As mentioned above,  the presence of gravity in the latter experiments induces a strong difference between the two works. \cite{Salami_EJECE_2017} indeed first place their sample at the bottom plane and loading frame and then, the lateral contacts are added. 
\revP{Consequently, in contrast to our simulations, the lateral contacts are not equivalent to the top and bottom ones due to their different ways of application and the sample is pre-loaded along diameter joining the top and bottom contacts.}
It would have been interesting to experimentally remove the aforementioned pre-load by using an horizontal experimental set-up instead of a vertical one. 
To support this analysis we have slightly shifted the lateral walls (\textit{i.e.} those located at $\alpha=\pi/4$) to remove their initial contact with the sample and thus favour the stress propagation along the diameter joining the loading point and its diametrically opposed counterpart.
In doing so, the bottom contact is indeed preloaded before the lateral contacts become active and our configuration is then closer to the experiments of~\cite{Salami_EJECE_2017}.  
The shifts are expressed as a function of $\zeta_0$, the deformation at the force peak of the top contact observed for the four point configuration in which all the contacts are initially active.
In other words, $\zeta_0$ corresponds to the deformation for which the force at the top contact reported in figure~\ref{fig:patterns_4pts} is maximum.
We have tested values of the aforementioned shift  between $10\%$ and $159\%$ $\zeta_0$. 
It should be pointed out that, \revric{due to the quasi-brittle behavior of the numerical model}, corresponding distances are extremely small \textit{i.e.} lower than $0.1\%$ of the particle diameter. 
The smallest shift used leads to an initiation of a vertical crack. Yet such an initiation does not necessarily lead to the propagation of the crack upwards across the sample since it depends on the loading of the lateral contact walls. 
The shift between the lateral-contacts and the studied particle has indeed to be large enough (\textit{e.g.} larger or equal than $\zeta_0$) to observe a diametrical crack that is not only initiated but also propagates upwards across the sample in agreement with the experimental work of~\cite{Salami_EJECE_2017} as shown in figure~\ref{fig:forces_4points_cent}(a) and (b). 
\begin{figure}[htb]
\begin{center}
\includegraphics*[width=0.85\columnwidth]{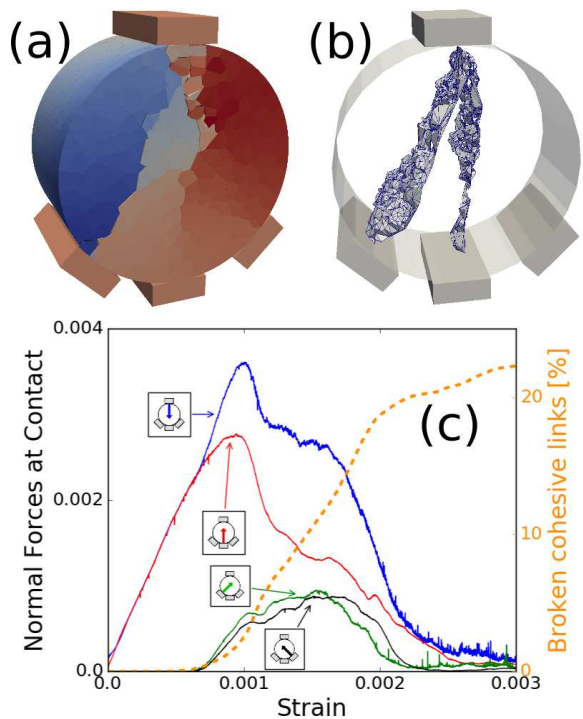}
\caption{Displacement field for the numerical simulations of the four contact configuration in which only the top and bottom contacts are initially active (a) and the corresponding fracture surface (b). Evolution with strain of (i) the number of broken cohesive links and (ii) of the normal forces at contacts in the case of elastic interactions \textit{i.e.} case A (c). Initially the lateral walls are not in contact with studied sample but shifted from a distance equal to the deformation corresponding to the peak of the normal force 
at the top contact 
when the four contact are initially active  (Fig.~\ref{fig:patterns_4pts}).}\label{fig:forces_4points_cent}
\end{center}
\end{figure}
Figure~\ref{fig:forces_4points_cent}(c) depicts the forces at contacts and the percentage of broken cohesive links \textit{versus} strain for a shift of the lateral contacts equal to $\zeta_0$.
The force at the bottom wall is always larger than those at the lateral contacts explaining why, in such a case, a diametrical crack is observed. To be more precise, as long as the lateral 
walls are not in contact with the sample, due to force balance, the normal force at the bottom wall equals that of the loading point (top wall). When the lateral walls touch the sample, the aforementioned equality is no more verified, the normal forces at the lateral walls start to increase and the bottom normal force becomes lower than that of the top wall. 
Predicting if a crack propagates or not along a sample remains a complicated task since, as mentioned above, it is influenced by the \revric{initial position} to the lateral walls.

These results which highlight the importance of the initial conditions (\textit{i.e.} contact locations and pre-load) are of crucial importance to understand and optimize crushing processes. In other words, the crushing properties of a particle during the multi-contact test is not only influenced by the number of  contacts but also by the history of loading \textit{i.e.} the knowledge of the initial stresses. 
\revPbis{The consequences are important since the fracture pattern might change due to a slight shift of a contact edge. The observed results originate from  the quasi-brittle behavior of the particle.}
These results shed light on the need to design new experiments to study multicontacts crushing since the forces exerted by each contact must be carefully controlled.

%

\section{Conclusions}
{The effect of the contact orientation distribution on the crushing of granular materials was studied by means of experimental and numerical tools.
At first, a simple dynamic experiment was presented to highlight the effect of the location of three contact edges on the fracture pattern and the strength of a model cylindrical grain. The sample was placed on two bottom contact edges and received an impact by a falling weight at the top. The energy required to fracture the particle was shown to be an increasing function of the angle between the plane connecting a bottom contact edge to the centerline of the cylinder and a vertical plane. Peculiar crack patterns were also observed, connecting the impact edge to the other contact edges.\\
A discrete model of grain fracture~\cite{neveu2016} was then employed for a numerical analysis of the effect of contact aniso\-tropy. The cylindrical grain was discretized by means of a space filling {Vorono\"\i} tessellation, and submitted to a  compression test for different values of angle $\alpha$. 
In agreement with experiments, simulations predict a strong effect of the contact orientation on the strength of the particle as well as similar fracture patterns. In particular, a linear reversible elastic behavior was found for small displacements of the top wall, and a progressive damage was experienced for larger displacements. 
Our results also emphasize the importance of the initial stresses on the contact points on the fracture patterns. Yet, any study focused on multicontact crushing has to control precisely these latter quantities.
The proposed numerical model appears therefore a promising tool for understanding fracture mechanisms at microscale with the purpose of optimizing the crushing process.}
\revPbis{In summary, our experimental and numerical results show the importance of contacts location on the fracture pattern and the apparent strength of the particle.
Two typical types of fracture are observed: a diametrical and an inverted Y-shaped fracture both being likely to superimpose. They are strongly influenced by the properties of the contacts and, due the quasi-brittle behavior of the particle, any infinitesimal gap between a contact edge and the particle may modify the fracture pattern.} \\
Future work will deal with more complex grain shapes and configuration of contacts. 
\revP{Here we focused on homogeneous materials but our method can be directly applied to a wide range of materials as long as the internal structure of the grains is known.  To test the generality of our results, we will also (i) use more complex failure criteria to model other types of material and (ii) introduce defects in the grain structure and thus potentially induce size effects.}

\section*{Acknowledgements}
The numerical simulations were carried out at the CCIPL
(Centre de Calcul Intensif des Pays de la Loire) under the
project \textit{Simulation num\'erique discr\`ete de la fracture des mat\'eriaux granulaires}.

\section{References}
\bibliographystyle{abbrvnat}
\bibliography{biblio}

\end{document}